\documentclass{elsart}
\usepackage{epsfig}

\usepackage{amssymb}

\newcommand{\lsim}{\raisebox{-.4ex}{$\stackrel{<}{\scriptstyle \sim}$}}
\newcommand{\gsim}{\raisebox{-.4ex}{$\stackrel{>}{\scriptstyle \sim}$}}
\begin{document}

\begin{frontmatter}

\title{Variation of the
gas and radiation content in the sub-Keplerian accretion disk around black holes and its
impact to the solutions}

%\maketitle
\author{Banibrata Mukhopadhyay\thanksref{email1}},
\author{Parikshit Dutta$^2$}
\address{1. Astronomy and Astrophysics Programme, Department of Physics,
Indian Institute of Science, Bangalore 560012, India}
\address{2. Department of Physics, Indian Institute of Technology, Kanpur 208016, India
}

\thanks[email1]{bm@physics.iisc.ernet.in}

%\pagerange{\pageref{firstpage}--\pageref{lastpage}} \pubyear{2008}

%\maketitle
%\newpage
\begin{abstract}

We investigate the variation of the gas and the
radiation pressure in accretion disks during the infall of matter to the black hole and its
effect to the flow. While the flow far away from the
black hole might be non-relativistic,
in the vicinity of the black hole it is expected to be relativistic behaving more
like radiation. Therefore, the ratio of gas pressure to total pressure ($\beta$) and 
the underlying polytropic index ($\gamma$) should not be constant throughout the flow.
   We obtain that accretion flows
exhibit significant variation of $\beta$ and then $\gamma$,
which affects solutions described in the standard literature based on constant $\beta$.
Certain solutions for a particular set of initial parameters with a
constant $\beta$ do not exist when the variation of $\beta$ is incorporated appropriately.
   We model the viscous sub-Keplerian accretion disk with a nonzero component of
advection and pressure gradient around black holes
by preserving the conservations of mass, momentum, energy,
supplemented by the evolution of
$\beta$.
By solving the set of five coupled differential equations, we obtain the thermo-hydrodynamical
properties of the flow.
   We show that during infall, $\beta$ of the flow could vary upto $\sim 300\%$, while $\gamma$ upto $\sim 20\%$.
   This might have a significant impact to the disk solutions in explaining observed data,
e.g. super-luminal jets from disks, luminosity, and then extracting fundamental properties
from them. Hence any conclusion based on
constant $\gamma$ and $\beta$ should be taken with caution and
corrected.

\end{abstract}

\end{frontmatter}

%\maketitle
%\label{firstpage}

\section{Introduction}

Gas flows in the vicinity of a compact object, particularly black hole and neutron star,
are expected to be highly relativistic. However, far away of it, when infalling matter comes off, 
e.g., a companion star, flows are rather non-relativistic. Therefore, during 
accretion of matter from a large distance, 
flows should transit from the non-relativistic to relativistic regime while reaching the
vicinity of the black hole horizon or the neutron star
surface, when flow is expected to be sub-Keplerian in nature. Indeed, it is generally believed
that the velocity and temperature at the inner edge of accretion disk are very high.
%(e.g. Corbel et al. 2003). 
%which has been supported by the theory (e.g. Rajesh \& Mukhopadhyay
%2009, 2010) showing the electron and ion temperatures could be as high as, respectively, 
%$\sim 10^{9.5}$K and $10^{11.8}$K. However, far away the temperature could be $50$ times lower.

Not only the cases in accretion, highly relativistic flows are involved in astrophysical jets
from galactic, extragalactic black holes, gamma-ray bursts etc. (e.g. Zensus 1997;
Mirabel \& Rodr\'iguez 1999; M\'esz\'aros 2002), with the velocity $0.9-0.98$ times the
speed of light. The collimated dipolar outflows emerging from deep inside collapsars, according
to the collapsar model of gamma-ray burst (Woosley 1993), are expected to achieve 
a Lorentz factor more than $100$. As the accretion and outflow/jet are expected to be
coupled (influencing each other; Bhattacharya, Ghosh \& Mukhopadhyay 2010), any change/transition
(e.g. as mentioned above) in the accretion flow would influence the jet and hence the underlying inferences.

When the velocity varies from the non-relativistic to relativistic (ultra-relativistic) regime,
the corresponding polytropic index ($\gamma$) in the equation of state (EOS), i.e. the ratio of
gas to total pressure ($\beta$) in the present context, should
not remain constant. Note that $\gamma$ and $\beta$ can be approximated as constant only if the
flow remains non-relativistic or ultra-relativistic throughout.
Therefore, the existing solutions for such flows with a fixed $\gamma$ need
to be verified and corrected if necessary with a relativistically correct EOS (see e.g. Chandrasekhar 1938).

Most of the studies of accretion disks around compact objects 
have approximated $\gamma$ to be constant throughout the flow, which cannot be the correct
description for the reasons described above. Blumenthal \& Mathews (1976) introduced, for
the first time to best of our knowledge, variable
$\gamma$ and proposed an appropriate EOS in obtaining solutions for spherical accretion
and wind around Schwarzschild black holes. Recently, Meliani et al. (2004) used their
EOS and obtained the general solutions for the relativistic Parker winds incorporating varying $\gamma$
according to the nature of the flow. Ryu et al. (2006) proposed a new EOS suitable 
for non-relativistic and relativistic regimes with varying $\gamma$, to implement
in the simulations of relativistic hydrodynamics. Very recently, Chattopadhyay \& Ryu (2009) 
showed that composition of the fluid plays an important role to determine the value of $\gamma$
and then solutions of inviscid spherical flows around black holes. This would have impact
on a hot accretion disk as well which was shown to exhibit significant nuclear burning, 
producing the variation of compositions in the disk (Chakrabarti \& Mukhopadhyay 1999,
Mukhopadhyay \& Chakrabarti 2000).
Earlier, while Chen \& Taam (1993) discussed structure and stability of the transonic optically thick 
accretion disks without restricting the gas and radiation content therein, did not show 
the actual variation of the gas/radiation content as a function of radial coordinate and then its impact to 
the solutions.

The present work introduces a self-consistent variation of $\gamma$ 
in solving viscous flows in the accretion disk around black holes.  
We essentially concentrate on the region of the flows which is expected to be
sub-Keplerian in nature. Therefore, following previous work (Muchotrzeb \& Paczynski 1982;
Abramowicz et al. 1988; Chakrabarti 1996; Mukhopadhyay \& Ghosh 2003; Rajesh \& Mukhopadhyay 2010a)
we model the flow from the Keplerian to
sub-Keplerian transition radius ($x_t$) to the black hole event horizon ($x_+$). We plan to
concentrate, in particular, the class of flows which is not geometrically very thick, yet
sub-Keplerian in nature including a nonzero component of advection and gradient of pressure
in general. Hence, the underlying equations can be integrated/averaged vertically without losing important physics, 
in obtaining the solutions. 
Therefore, following previous work (Muchotrzeb \& Paczynski 1982;
Abramowicz et al. 1988; Chakrabarti 1996; Mukhopadhyay \& Ghosh 2003; Rajesh \& Mukhopadhyay 2010a, 2010b), 
we consider vertically integrated
infall consisting of gas and radiation. While the flow around 
$x_t$ might be non-relativistic with a high $\gamma$ (depending on the 
size of the disk), as it advances towards $x_+$, $\gamma$ must be varying.
We plan to investigate, how the self-consistent variation of $\gamma$ over the disk radii  
affects the established solutions with a fixed $\gamma$. 

In the next section we establish the model equations describing flows. Subsequently, we discuss solutions 
in \S3 and finally summarize the results in \S4.

%__________________________________________________________________

\section{Model Equations}

The basic equations describing conservations of mass and momentum in 
the vertically integrated disk remain
same as of earlier works which assumed $\gamma$ and then $\beta$ 
to be constant throughout (e.g. Abramowicz et al. 1988; Chakrabarti 1996; Rajesh \& Mukhopadhyay 2010b).
Abramowicz et al. (1988) also showed that the geometrically slimmer flow is stable for $\beta >0.4$.
For simplicity, following e.g. Narayan \& Yi (1994) and Chakrabarti (1996), we assume
that the heat radiated out to be proportional to the heat generated by viscous dissipation.
Hence, following the same authors, we define a parameter $f$ determining the fraction
of matter advected in. Therefore, we recall the equation describing conservation
of energy in the vertically integrated disk from Chakrabarti (1996) and Mukhopadhyay \& Ghosh (2003).
However, in reality, $f$ should be determined, rather
fixing, selfconsistently as done by Rajesh \& Mukhopadhyay (2010a) very recently for hot 
optically thin flows around black holes. 
%Moreover, the hot flows might appear optically thin (as well as geometrically
%thick) such that the bremsstrahlung and synchrotron components of radiation can not be 
%neglected (Rajesh \& Mukhopadhyay 2009, 2010). In future work, we would like to include such
%physics in the present context, particularly in studying very hot flows, 
%once the present results confirm the work worth pursuing.

For the present purpose, all the above mentioned equations of conservation are supplemented by 
an equation describing
the variation of $\beta$ as a function of radial coordinate, assuming a single temperature fluid.
Earlier authors (e.g. Chen, Abramowicz \& Lasota 1997) already formulated single temperature optically thin 
disk flows including bremsstrahlung cooling for constant $\beta$.
Therefore, without repeating the description of equations of conservation, 
assuming the flow to be a mixture of perfect gas and radiation (Abramowicz et al. 1988;
Chakrabarti 1996; Rajesh \& Mukhopadhyay 2010a,b), we recall straight away
\begin{eqnarray}
%\nonumber
&&\beta=\frac{P_{\rm gas}}{P}\,\,=\frac{P_{\rm gas}}{P_{\rm rad}+P_{\rm gas}},\,\,
P_{\rm gas}=\frac{\rho k T}{\mu_i m_p},
\label{bet1}
\end{eqnarray}
where $k$ is the Boltzmann constant, $m_p$ the mass
of a proton, $\mu_i$ the mean molecular weight of ions, and $P$, $\rho$, $T$  
are respectively the total pressure, density, temperature in $K$ of the flow.
Now for an optically thick flow, as was initiated in the limit of geometrically slim disk
by Muchotrzeb \& Paczynski (1982), Abramowicz et al. (1988),
\begin{eqnarray}
P_{\rm rad}=P_{\rm bb}={\rm pressure\,\,from\,\,blackbody\,\,radiation}=\frac{a T^4}{3},
\label{prbb}
\end{eqnarray}
where $a$ is the radiation constant.
On the other hand, for the optically thin hot flows (e.g. Narayan \& Yi 1995; Mandal \& Chakrabarti 2005;
Rajesh \& Mukhopadhyay 2010a,b) in the presence of bremsstrahlung as radiation mechanism, 
which are presumably geometrically thicker
than that of Muchotrzeb \& Paczynski (1982), Abramowicz et al. (1988),
\begin{eqnarray}
\nonumber
P_{\rm rad}=P_{\rm br}&=&{\rm pressure\,\,from\,\,bremsstrahlung\,\,radiation}\\
%&=&1.4\times 10^{-27}\frac{\rho^2}
%{m_p^2\mu_e\mu_i}\,h\,T_e^{1/2}\,(1+4.4\times 10^{-10} T_e)\,\frac{4GM}{3c^3},
%&\sim &5\times 10^{-66}\frac{\rho^2}
&\sim &2\times 10^{-18}\,\rho^2\,
h\,T^{1/2}\,(1+4.4\times 10^{-10} T)\,M
%\beta=\frac{\frac{\rho k T}{\mu m_p}}{\frac{\sigma T^4}{3}+1.67\times 10^{10}\rho^2\,hT^{1/2}+\frac{\rho k T}{\mu m_p}},
\label{prbr}
\end{eqnarray}
in CGS unit, where $h$ is the half-thickness of the flow in units of $GM/c^2$, 
$G$ the Newton's gravitation constant, $M$ the mass of the black hole, $c$ the speed of light. 
%$\mu_e$ the mean molecular weight of electrons.
Note that, in the first approximation (Rajesh \& Mukhopadhyay 2010b), 
we do not consider other radiation mechanisms which could be effective in
optically thin flows.
In a future work, such effects e.g. synchrotron radiation, inverse-Comptonization 
could be included, once the present results confirm such a study worth pursuing.
With the choice of $\gamma P=\rho c_s^2$ (Mukhopadhyay \& Ghosh 2003), we further
write
\begin{eqnarray}
c_s^2=\frac{8-3\beta}{6-3\beta}\frac{k T}{\mu m_p \beta},
\label{bet2}
\end{eqnarray}
where $c_s$ is the sound speed of the flow.

\subsection{Optically thick flows}

First we set up the model equations for optical thick and geometrically slim flows.
Eliminating $T$ from eqns. (\ref{bet1}), (\ref{prbb}) and (\ref{bet2}) we obtain
\begin{eqnarray}
R\frac{d\beta}{dx}+\frac{6(1-\beta)}{c_s}\frac{dc_s}{dx}+(\beta-1)\frac{1}{\rho}\frac{d\rho}{dx}=0,
\label{betd}
\end{eqnarray}
where 
\begin{eqnarray}
R=\frac{1}{\beta}+3(1-\beta)\left(\frac{1}{\beta}-\frac{6}{(8-3\beta)(6-3\beta)}\right)
\label{betcof}
\end{eqnarray}
and $x$ is the radial coordinate in units of $GM/c^2$. As $\beta$ and $\gamma$
are related by (Narayan \& Yi 1995; Ghosh \& Mukhopadhyay 2009)
\begin{eqnarray}
\beta=\frac{6\gamma-8}{3(\gamma-1)},
\label{betgam}
\end{eqnarray}
from eqn. (\ref{betcof}) one can easily obtain $d\gamma/dx$ as well.
Now replacing $P$, using eqns. (\ref{bet1}) and (\ref{bet2}), from the momentum and energy conservation equations 
and then eliminating $\rho$ using the mass conservation equation (see Mukhopadhyay \& Ghosh
2003, for details), we obtain 
\begin{eqnarray}
\left(\frac{\gamma\vartheta}{K c_s^2}-\frac{1}{\vartheta}\right)\frac{d\vartheta}{dx}+\frac{1}{c_s}
\left(\frac{L}{K}-1\right)\frac{dc_s}{dx}=\frac{3}{2x}+\frac{\gamma}{K c_s^2}\left(\frac{\lambda^2}{x^3}
-F\right)-\frac{1}{2F}\frac{dF}{dx},
\label{mom1}
\end{eqnarray}
\begin{eqnarray}
A\frac{d\vartheta}{dx}+\vartheta B\frac{dc_s}{dx}+\frac{f I_n Z}{x}\frac{d\lambda}{dx}=\frac{2 f I_n Z \lambda}
{x^2}-\frac{3}{2}\frac{A\vartheta}{x}+\frac{A\vartheta}{2F}\frac{dF}{dx},
\label{mom2}
\end{eqnarray}
\begin{eqnarray}
\nonumber
&&\frac{x}{\vartheta}\left[K\,Z-(1+K)\alpha\vartheta^2\right]\frac{d\vartheta}{dx}-\frac{x}{c_s}(L-K+1)(Z-\alpha\vartheta^2)
\frac{dc_s}{dx}+\vartheta\frac{d\lambda}{dx}\\&&=\left(\frac{5}{2}-\frac{3}{2}K\right)Z+(K-1)\left(\frac{3}{2}
\alpha\vartheta^2+\alpha x\frac{I_{n+1}}{I_n}\frac{c_s^2}{\gamma}\frac{1}{2F}\frac{dF}{dx}\right),
\label{mom3}
\end{eqnarray}
where $\vartheta$ and $\lambda$ are the radial velocity and specific angular momentum of the disk 
respectively, $f$ is kept
constant for a particular flow (Mukhopadhyay \& Ghosh 2003; Rajesh \& Mukhopadhyay 2010b);
$f\rightarrow 1$ corresponds to the advection of entire energy in and $f\rightarrow 0$ to
no advection, $\alpha$
the viscosity parameter, $F$ the pseudo-Newtonian 
gravitational force due to a rotating black hole (Mukhopadhyay 2002), the functions $A,B,Z,K,L$ are
given by
\begin{eqnarray}
\nonumber
A=\frac{c_s^2}{\gamma}\frac{\Gamma_1-K}{\Gamma_3-1},\,\,\,B=\frac{c_s}{\gamma}\frac{L-K+\Gamma_1}{\Gamma_3-1},
\,\,\,Z=\alpha\left(\frac{I_{n+1}}{I_n}\frac{c_s^2}{\gamma}+\vartheta^2\right),\\
K=1+\frac{6(\beta-1)}{(6-3\beta)(8-3\beta)R},\,\,\, L=2+\frac{36(1-\beta)}{R(6-3\beta)(8-3\beta)},
\label{mcof}
\end{eqnarray}
where $I_n, I_{n+1}, \Gamma_1, \Gamma_3$ have their usual meaning, discussed by the previous authors
in detail (e.g. Mukhopadhyay \& Ghosh 2003; Rajesh \& Mukhopadhyay 2010a,b). 
%Note that unlike the
%previous works, here $n$ [$=1/(\gamma-1)$] and then $I_n,I_{n+1}$ no longer remain constant. 

Now combining eqns. (\ref{mom1}), (\ref{mom2}), (\ref{mom3}) we obtain
\begin{eqnarray}
\frac{d\vartheta}{dx}=\frac{f_1}{f_2},\,\,\,{\rm where}
\label{dvdx}
\end{eqnarray}
\begin{eqnarray}
\nonumber
f_1=\left[\frac{3}{2x}+\frac{\gamma}{K\,c_s^2}\left(\frac{\lambda^2}{x^3}-F\right)-\frac{1}{2F}
\frac{dF}{dx}\right]\left[\vartheta^2 B+\frac{f I_n Z}{c_s}(L-K+1)\right.\\
\nonumber
\left.(Z-\alpha^2\vartheta^2)\right]-\frac{1}{c_s}\left(\frac{L}{K}-1\right)\left[\vartheta\left(\frac{2f I_n 
Z\lambda}{x^2}-\frac{3}{2}\frac{A\vartheta}{x}+\frac{A\vartheta}{2F}\frac{dF}{dx}\right)\right.\\
\left. -\frac{f I_n Z}{x}\left(\left(\frac{5}{2}-\frac{3}{2} K\right)Z+(K-1)\left(\frac{3}{2}\alpha\vartheta^2+
x\alpha\frac{I_{n+1}}{I_n}\frac{c_s^2}{\gamma}\frac{1}{2F}\frac{dF}{dx}\right)\right)\right],
\label{dvdx1}
\end{eqnarray}
\begin{eqnarray}
\nonumber
f_2=\left(\frac{\gamma\vartheta}{K\,c_s^2}-\frac{1}{\vartheta}\right)\left(\vartheta^2B+(L-K+1)(Z-
\alpha\vartheta^2)\frac{fI_nZ}{c_s}\right)\\-\frac{1}{c_s}\left(\frac{L}{K}-1\right)
\left(A\vartheta-\frac{KZ-(1+K)\alpha\vartheta^2}{\vartheta}fI_nZ\right).
\label{dvdx2}
\end{eqnarray}
We strictly follow the methodology adopted by Rajesh \& Mukhopadhyay (2010a) in solving eqn. (\ref{dvdx}).
Then following Mukhopadhyay \& Ghosh (2003) and Rajesh \& Mukhopadhyay (2010a,b)
the outer boundary $x_t$, where the flow becomes purely sub-Keplerian, is specified with the condition 
$\lambda/\lambda_K=1$ ($\lambda_K$ being value of Keplerian $\lambda$). As matter advances from $x_t$ to
$x_+$, it reaches the critical radius $x=x_c$ when $f_1=f_2=0$. Hence from $f_2=0$, the algebraic 
equation of Mach number 
($M_{ac}$) at $x_c$ is given by
\begin{eqnarray}
{\cal A}\,M_{ac}^4+{\cal B}\,M_{ac}^2+{\cal C}=0,\,\,\,{\rm where}
\label{mac}
\end{eqnarray}
\begin{eqnarray}
\nonumber
&&{\cal A}=\frac{L-K+\Gamma_1}{K(\Gamma_3-1)}+\frac{L-K+1}{K}\alpha^2f I_{n+1}-\left(\frac{L}{K}-1\right)
\alpha^2fI_n,\\
\nonumber
&&{\cal B}=-\frac{1}{\gamma(\Gamma_3-1)}\left[\left(\frac{L}{K}-1\right)(\Gamma_1-K)+(L-K+\Gamma_1)\right]\\
\nonumber
&&+\frac{(L-K+1)}{K}\frac{\alpha^2fI_{n+1}}{\gamma}\frac{I_{n+1}}{I_n}-(L-K+1)\alpha^2fI_{n+1}\\
\nonumber
&&+\left(\frac{L}{K}-1\right)\alpha^2fI_n\frac{K-1}{\gamma}\frac{I_{n+1}}{I_n},\\
&&{\cal C}=-\frac{(L-K+1)}{\gamma^2}\alpha^2fI_{n+1}\frac{I_{n+1}}{I_n}+\left(\frac{L}{K}-1\right)\alpha^2
fI_nK\left(\frac{I_{n+1}}{I_n\gamma}\right)^2.
\label{mac1}
\end{eqnarray}
Then $M_{ac}$ can be easily computed from eqn. (\ref{mac}) for given $\alpha$ and $f$, once 
$\gamma$ and therefore $\beta$ is known at $x_c$, say $\gamma_c$ and therefore $\beta_c$.
Subsequently, $c_s$ at $x_c$, say $c_{sc}$, can be
obtained from $f_1=0$, once $x_c$ and the specific angular momentum of the disk at this radius 
$\lambda_c$ are obtained.

Now using the conservation of mass, eliminating $d\rho/dx$ from eqn. (\ref{betd}), we obtain
\begin{eqnarray}
%\nonumber
\frac{d\beta}{dx}=\frac{1}{R}\left[-\frac{7(1-\beta)}{c_s}\frac{dc_s}{dx}-\frac{(1-\beta)}{\vartheta}
\frac{d\vartheta}{dx}+(1-\beta)\left(-\frac{3}{2x}+\frac{1}{2F}\frac{dF}{dx}\right)\right].
\label{betdf}
\end{eqnarray}
Hence, by solving eqns. (\ref{dvdx}) and (\ref{betdf}), along with any two of eqns. 
(\ref{mom1}), (\ref{mom2}), (\ref{mom3})
simultaneously, we obtain $\vartheta, c_s, \lambda, \beta$ (and then $\gamma)$ as functions of $x$ from
$x_t$ to $x_+$.

\subsection{Optically thin flows}

In presence of bremsstrahlung radiation for optically thin flows, from 
eqns. (\ref{bet1}), (\ref{prbr}) and (\ref{bet2}) we obtain
\begin{eqnarray}
%R\frac{d\beta}{dx}+\left(\frac{W_2}{c_s}-\frac{2W_4 T}{c_s}\right)\frac{dc_s}{dx}+\frac{W_1}{\rho}\frac{d\rho}{dx}=W_3,
R\frac{d\beta}{dx}+\frac{(1-\beta)}{c_s}\frac{dc_s}{dx}+\frac{(1-\beta)}{\rho}\frac{d\rho}{dx}=W,
\label{bettn}
\end{eqnarray}
where 
\begin{eqnarray}
%\nonumber
%&&R=\frac{1}{\beta}-T\,W_4\,W_5,\,\,\,W_1=1-\beta,\,\,\,W_2=2(1-\beta),\,\,\,W_3=\frac{1-\beta}{2}\left(\frac{1}{F}\frac{dF}{dx}-
%\frac{1}{x}\right),\,\,\,W_4=\frac{1-\beta}{2T},\\
R=\frac{1}{\beta}-\frac{1-\beta}{2}\left(\frac{1}{\beta}-\frac{6}{(8-3\beta)(6-3\beta)}\right),\,\,\,\,
W=\frac{1-\beta}{2}\left(\frac{1}{F}\frac{dF}{dx}- \frac{1}{x}\right).
\label{betcof2}
\end{eqnarray}
The remaining equations are same as in an optically thick flow, with $R$ given by eqn. (\ref{betcof2}).
Subsequently, as in \S 2.1, eliminating $d\rho/dx$ from eqn. (\ref{bettn}), we can obtain $d\beta/dx$
in terms of $d\vartheta/dx$ and $dc_s/dx$ and hence solve for 
$\vartheta, c_s, \lambda, \beta$ (and then $\gamma)$.

%__________________________________________________________________

\section{Solution}
First we discuss the solution for the optically thick flows which are
geometrically slimmer. Subsequently, we address the solution for the optically
thin and geometrically thicker flows.

\subsection{Optically thick flows}

We consider typical cases of accretion flow with $\beta$ at $x_t$ being $0\lsim\beta_t\lsim 1$;
extreme radiation dominated to extreme gas dominated flows at $x_t$.
Note that the general criterion for thermal stability of the flow
is $\beta >0.4$ (Abramowicz et al. 1988). 
We also assume that a part ($f$) of the energy dissipated in the 
infalling matter advects in.\footnote{$Q_{\rm avd}=
Q^+-Q^-=f\,Q^+$, when $Q_{\rm avd}$, $Q^+$ and $Q^-$ respectively the amount of energy advected in, 
dissipated and radiated out.}. However, note importantly that the value of $f$ is
not arbitrary, rather should be chosen according to the nature of the flow; whether it is 
gas dominated or radiation dominated, optically thin or thick. 
The flow consisting of significant radiation component
should be radiatively more efficient and hence should have a small $f$ compared to
a flow with a lesser content of radiation. 
%However, if the flow viscosity decreases, infall time scale will increase and hence $f$ might
%be lower evenif the flow consists of a significant fraction of gas.
However, the flow with significant radiation 
around a rotating black hole might have a larger $f$ compared to
its non-rotating counter part, as the radial velocity of infall matter is higher
for a rotating black hole with a shorter infall time scale.

\begin{figure*}
   \centering
   \includegraphics[width=9.8cm]{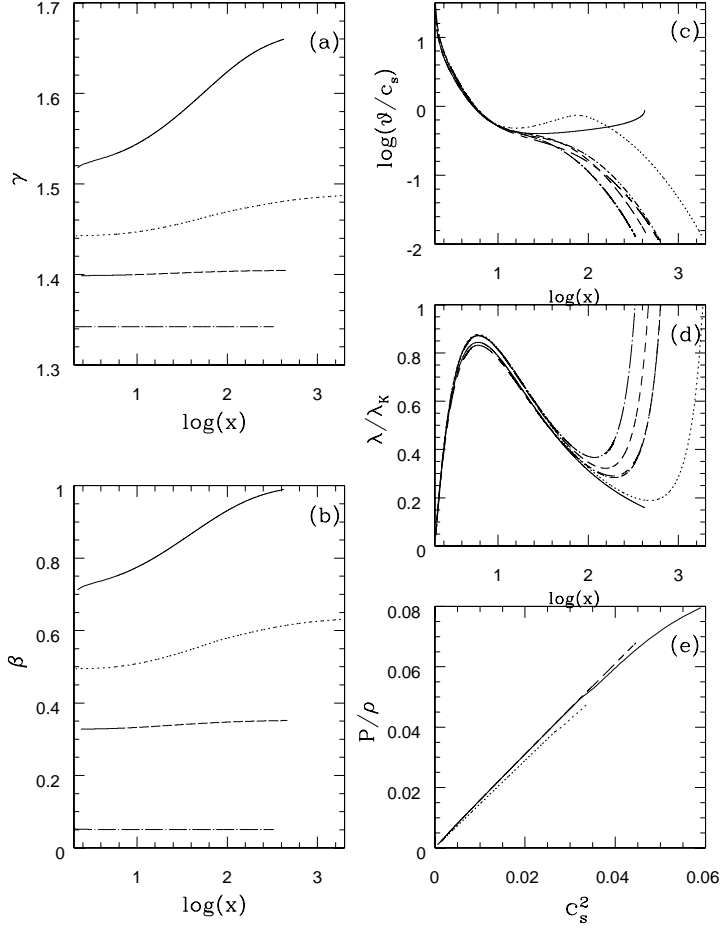}
   \caption{\small Variation of (a) polytropic index,
(b) ratio of gas pressure to total pressure, 
(c) Mach number, (d) ratio of specific angular momentum to 
the corresponding Keplerian value. In (a), (b), (c), (d) the solid, dotted, dashed, dot-long-dashed lines
corresponding to the different $\beta$ at transition radius are for $\beta_c=0.75,0.5,0.33,0.05$ respectively. 
In (c) and (d) long-short-dashed and dot-dashed lines
correspond to $\beta_c=0.75$, but respectively with a smaller $\lambda_c$ ($=3.05$) than that of solid line
($\lambda_c=3.1$) and with a constant $\gamma$ with $\lambda_c=3.1$. 
(e) The density-pressure-sound speed relation (EOS), when
the profiles shown by solid and dotted lines are the corresponding EOS of the solutions shown 
in (c) and (d) by solid and dotted lines
respectively, while dashed line corresponds to a case with smaller $\lambda_c$ than that of solid line.
Other parameters are $\alpha=0.01$, $f=0.5$, $x_c=5.8$, $a=0$, $M=10$. See Table 1 for detail.
}
              \label{figbeta}%
    \end{figure*}

Figures \ref{figbeta}a,b show how $\gamma$ and $\beta$ decrease as the flow advances from $x_t$ to $x_+$.
When $\beta_t\sim 1$, the flow is extremely gas dominated at $x_t$ and $\gamma_t\sim 5/3$. However, as
the flow advances in, the gas flow starts becoming relativistic, rendering the
increase of apparent content of radiation which decreases $\beta$. At $x_+$, $\beta$ decreases by $\sim 30\%$, 
while $\gamma$ by $\sim 9\%$. However, as $\beta_t$ and then $\gamma_t$ decreases, their variation
decreases as well. For $\beta_t\sim 0.33$, its variation during infall is only $\sim 7\%$, while for
a smaller $\beta_t$ variation is negligible. Therefore, for a gas dominated flow, the solutions of
accretion flows, particularly for vertically integrated geometrically thin/slim disks,
with a constant $\beta$ and then $\gamma$ as discussed in the existing literature, 
appear to be incorrect which need to be revised.

Figures \ref{figbeta}c,d show how the variation of $\beta$ affects the disk thermo-hydrodynamics,
particularly for gas dominated flows. In general, as $\beta$ decreases at a large distance from the black hole,
the content of radiation in the flow increases which leads to a smaller $\vartheta$ therein.
This is because, the increase of radiation makes the flow to be radiatively more efficient, rendering
the disk to be geometrically thinner and then more centrifugally dominated. Hence, ignoring the variation 
of $\beta$, when $\beta_t$ and $\beta_c$ are high, 
would make the disk to be gas pressure dominated geometrically thick and quasi-spherical 
throughout which could reflect a wrong picture. However, at the inner region of the disk,
$\vartheta$ profiles for all $\beta$ merge. This is because the flow therein is practically  
controlled by the black hole's gravity. Interestingly, while the flow with 
a constant $\beta$ ($=0.75$) would have a physical solution (dot-dashed line) for $\lambda_c=3.1$,
the variation of $\beta$ (with $\beta_c=0.75$), when other initial parameters remain unchanged, 
leads to an unphysical solution (solid line) 
with an O-type critical point which does not continue from $x_t$ to $x_+$. However, if $\lambda_c$
decreases slightly (to , e.g., $3.05$), then the flow with varying $\beta$ 
again exhibits a physical solution (long-short-dashed line). Figure \ref{figbeta}e shows how the
relation between $P$, $\rho$ and $c_s$, i.e. EOS, changes with the variation of $\beta$ in the
flow. Clearly $dP/d\rho\sim P/\rho$ (which is practically true for the present purposes)
does not vary linearly with $c_s^2$, unlike the cases modeled in the existing
literature based on constant $\gamma$.

\begin{figure*}
   \centering
   \includegraphics[width=10.5cm]{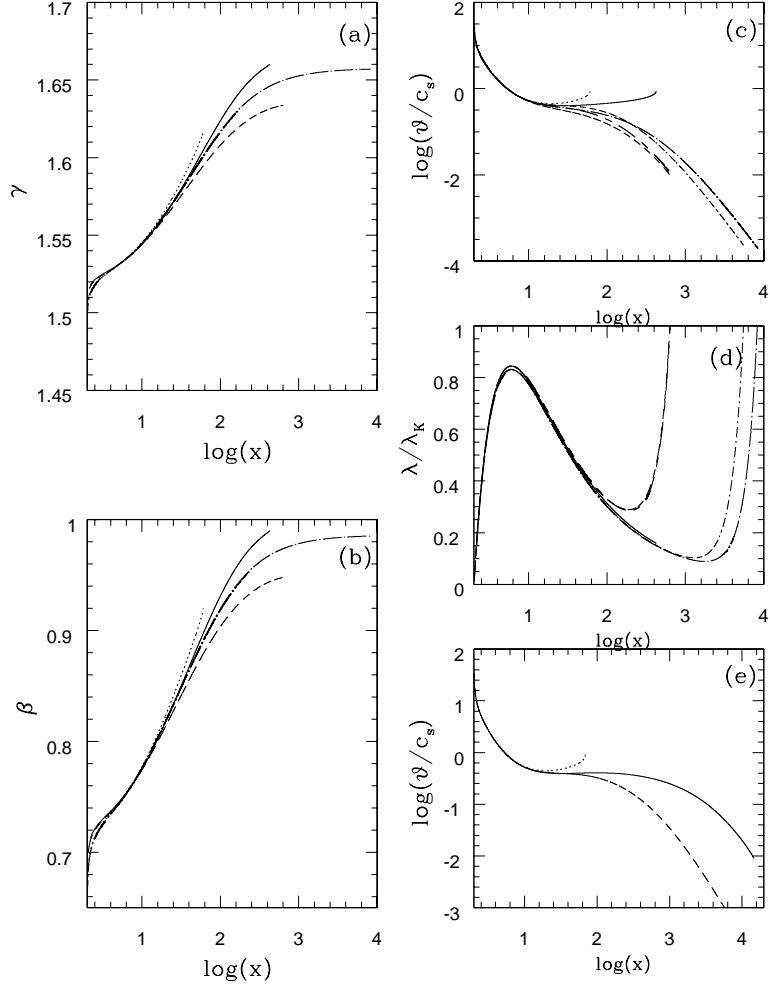}
   \caption{\small Variation of (a) polytropic index,
(b) ratio of gas pressure to total pressure, 
(c) Mach number, (d) ratio of specific angular momentum to 
the corresponding Keplerian value, as functions
of radial coordinate. In (a), (b), (c), (d) the solid and dotted lines are for
$\alpha=0.01$ and $0.0001$ respectively when $\lambda_c=3.1$, and dashed and dot-long-dashed lines
for $\alpha=0.01$ and $0.0001$ respectively when $\lambda_c=3.05$. 
In (c) and (d) long-short-dashed and dot-dashed lines
are corresponding solutions with constant $\gamma$ ($\sim 1.533$) for the cases of $\alpha=0.01$ and 
$\alpha=0.0001$ respectively when $\lambda_c=3.1$. 
(e) Mach number as a function of radial coordinate, when solid and dotted lines
are for $\alpha=0.01$ and $0.0001$ respectively where $\beta_c=0.745$, and dashed
line is for $\alpha=0.0001$ where $\beta_c=0.7$.
Other parameters are $f=0.5$, $x_c=5.8$, $a=0$, $M=10$. See Table 1 for detail.
}
              \label{figalfa}%
    \end{figure*}

In Fig. \ref{figalfa} we compare the change of variation of $\beta$, $\gamma$ and then 
corresponding $\vartheta$, with the change of $\alpha$. It is found that the effect of
variation of $\beta$ on disk thermo-hydrodynamics is stronger for a lower viscosity.
If $\beta_c$ is fixed, then $\beta$ is higher and its variation 
is steeper for a lower $\alpha$ at a large distance. Higher the $\beta$, higher the 
$\gamma$ in the flow is, inducing a quasi-spherical gas flow with a faster radial infall which 
may lead to an unstable/unphysical flow solution at a low $\alpha$. In other words,
a lower $\alpha$ corresponds to a lower rate of energy-momentum transfer in the disk. Hence,
to describe a steady infall, $\beta$ has to be larger, particularly at a large $x$, 
which implies a quasi-spherical
faster inflow of gas. Therefore, the assumption of a constant $\beta$ and then $\gamma$ ($\sim 1.533$)
would reflect an incorrect
solution of the disk flow, particularly for a gas dominated flow. Figures \ref{figalfa}c,d
show that physically interesting solutions, exhibiting flows coming from $x_t$ 
to $x_+$, require a smaller $\lambda_c$ than that considered in the literature based on a constant $\beta$. 
This supports quasi-spherical flows.
Figure \ref{figalfa}e compares solutions with different sets of $\beta_c$ and $\alpha$, and shows 
that the physical solution is achieved for a smaller $\beta_c$, when $\alpha$ is lower.

\begin{figure*}
   \centering
   \includegraphics[width=9.9cm]{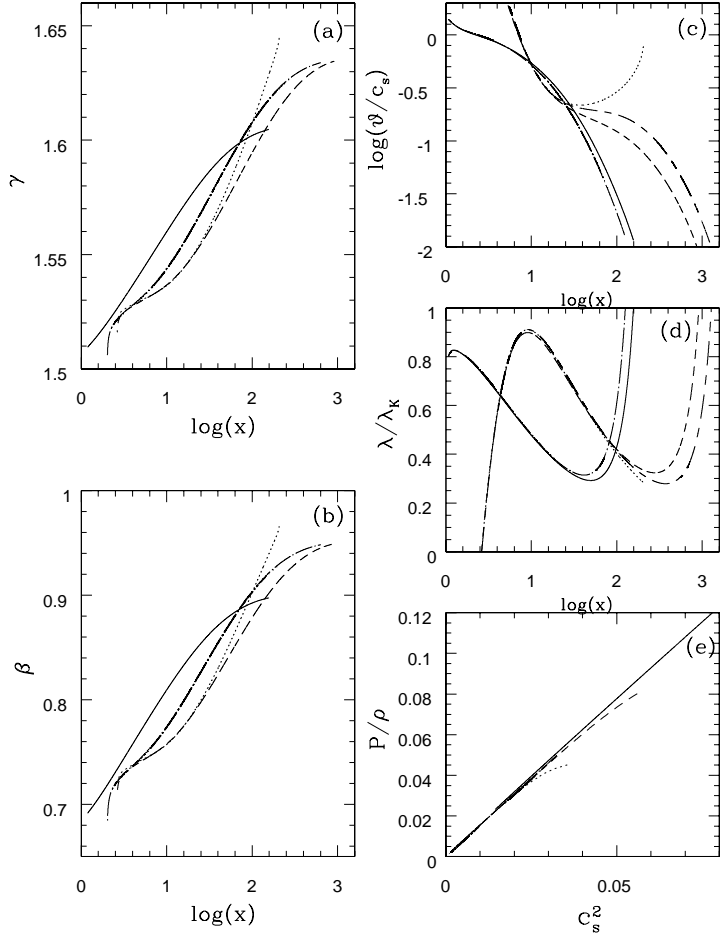}
   \caption{\small Variation of (a) polytropic index,
(b) ratio of gas pressure to total pressure, 
(c) Mach number, (d) ratio of specific angular momentum to 
the corresponding Keplerian value. In (a), (b) the solid, dotted, dashed, dot-long-dashed lines
respectively correspond to parameter sets $a=0.998, \lambda_c=1.7$; $a=-0.998, \lambda_c=4$; 
$a=-0.998, \lambda_c=3.95$; $a=0, \lambda_c=3.05$. In (c), (d) the solid, dotted, dashed, 
dot-long-dashed, dashed-long-dashed lines respectively correspond to parameter sets 
$a=0.998, \lambda_c=1.7$; $a=-0.998, \lambda_c=4$; $a=-0.998, \lambda_c=3.95$;
$a=0.998, \lambda_c=1.7$, with constant $\beta=0.75$; $a=-0.998, \lambda_c=4$, with constant 
$\beta=0.75$. The profiles shown by solid, dotted, dashed, dashed-long-dashed 
lines in (e) are
the corresponding density-pressure-sound speed relation (EOS)  
of the solutions shown by the same lines in (c) and (d).
Other parameters are $\alpha=0.01$, $f=0.5$, $M=10$. See Table 1 for detail.
}
              \label{figa}%
    \end{figure*}

From Fig. \ref{figa} we understand how the variation of Kerr parameter affects 
the solutions with
varying $\beta$. First we consider the maximum possible $\lambda_c$ obtained 
with constant $\beta$ (Rajesh \& Mukhopadhyay 2010a) for the 
respective cases of $a$. Interestingly,
while the physical solution around a co-rotating (prograde) black hole is available (solid line)
with varying $\beta$ for the same set of parameters as of constant $\beta$, $\lambda_c$ has to be smaller 
for a counter-rotating (retrograde) black hole (dotted and dashed lines). 
This is because a high $\beta_t$ renders a
quasi-spherical gas flow, which requires a smaller $\lambda$
allowing the matter to fall inwards. Note that while the maximum possible $\lambda_c$ for a co-rotating black hole
itself is smaller, for a counter-rotating black hole it is much larger. Hence, the allowed range of
$\lambda_c$ is shrunk for a counter-rotating black hole compared to that inferred from the analysis
based on a constant $\beta$. For clarity, in Figs. \ref{figa}c,d we also show the $\vartheta$ and
$\lambda$ profiles for $a=0.998$ 
and $-0.998$ (respectively dot-long-dashed and dashed-long-dashed lines) with a constant 
$\beta$ for maximum possible values of $\lambda_c$. 
Figure \ref{figa}e shows how the EOS changes with the variation of $a$,
when $\beta$ varies in the flow. Clearly $P/\rho$ does not vary linearly with $c_s^2$, 
unlike the cases with constant $\gamma$. Note that dashed and dashed-long-dashed lines
practically overlap each other, implying that the solution with a constant $\beta$ is recovered
by decreasing $\lambda_c$ in the case of varying $\beta$.

\begin{figure*}
   \centering
   \includegraphics[width=14cm]{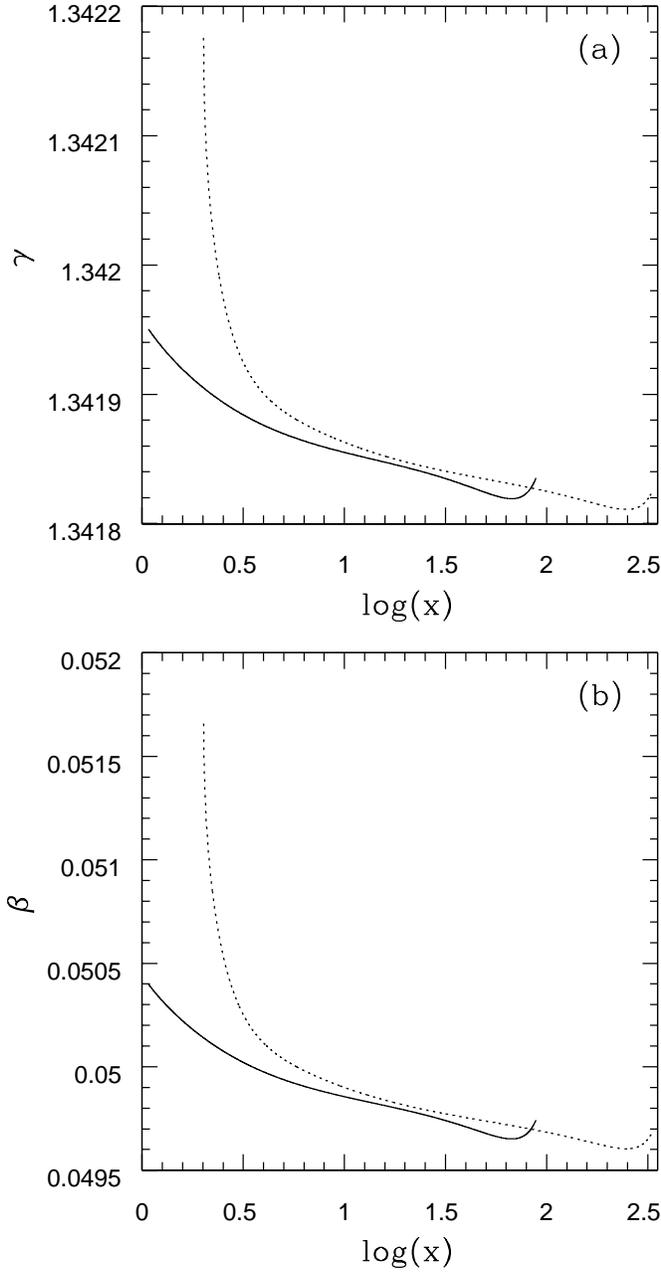}
   \caption{\small Variation of (a) polytropic index,
(b) ratio of gas pressure to total pressure, as functions of
of radial coordinate. The solid and dotted lines correspond to $a=0.998, \lambda_c=1.8$ 
and $a=0, \lambda_c=3.2$ respectively.
Other parameters are $\alpha=0.01$, $f=0.5$, $M=10$. See Table 1 for detail.
}
              \label{figrad}%
    \end{figure*}

We have already seen in Fig. \ref{figbeta} that the variation of $\beta$ is negligible for the radiation
dominated flows. This is because a small $\beta$ corresponds to a radiatively efficient 
geometrically thinner flow, which effectively corresponds to 
a relativistic flow of radiation throughout. Therefore, as the matter advances, the black hole's gravity 
has nothing to affect especially. However, close to the black hole, $\vartheta$ increases
noticeably and the corresponding infall time scale
of the matter decreases significantly, rendering a quasi-spherical flow which increases $\beta$ by $4\%$, as shown in Fig. \ref{figrad}.

\begin{figure*}
   \centering
   \includegraphics[width=14cm]{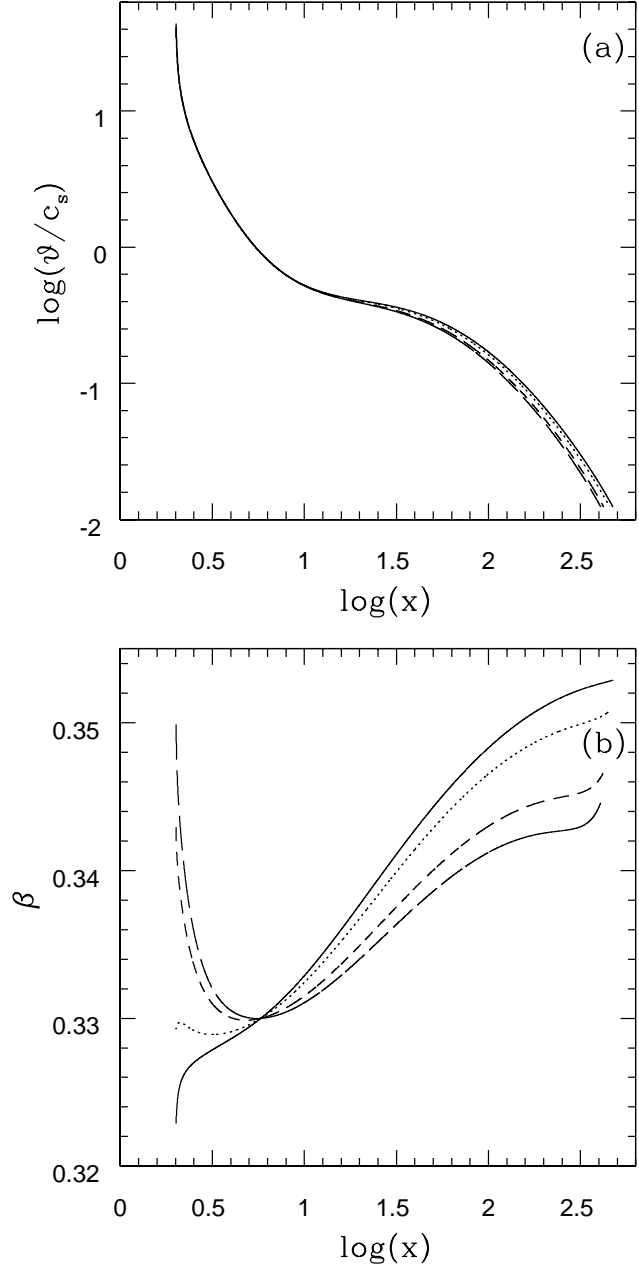}
   \caption{\small Variation of (a) Mach number, (b) ratio of gas pressure to total pressure, as functions of
of radial coordinate. The solid, dotted, dashed, long-dashed lines correspond to flows
with $f=0.1,0.3,0.5,0.9$ respectively, when $\beta_c=0.33$.
See Table 1 for detail.
}
              \label{figradcomf}%
    \end{figure*}

Figure \ref{figradcomf} shows that the variation of $f$ keeping other parameters
intact does not affect hydrodynamic properties, e.g. Mach number, significantly. 
However, as $f$ increases, the flow tends to become radiatively inefficient, rendering 
an increase in $\beta$ significantly close to the black hole. Therefore, any
(observed) property related to the inner accretion flow must be influenced by
$f$ and hence the variation of $\beta$.

\subsection{Optically thin flows}

\begin{figure*}
   \centering
   \includegraphics[width=14cm]{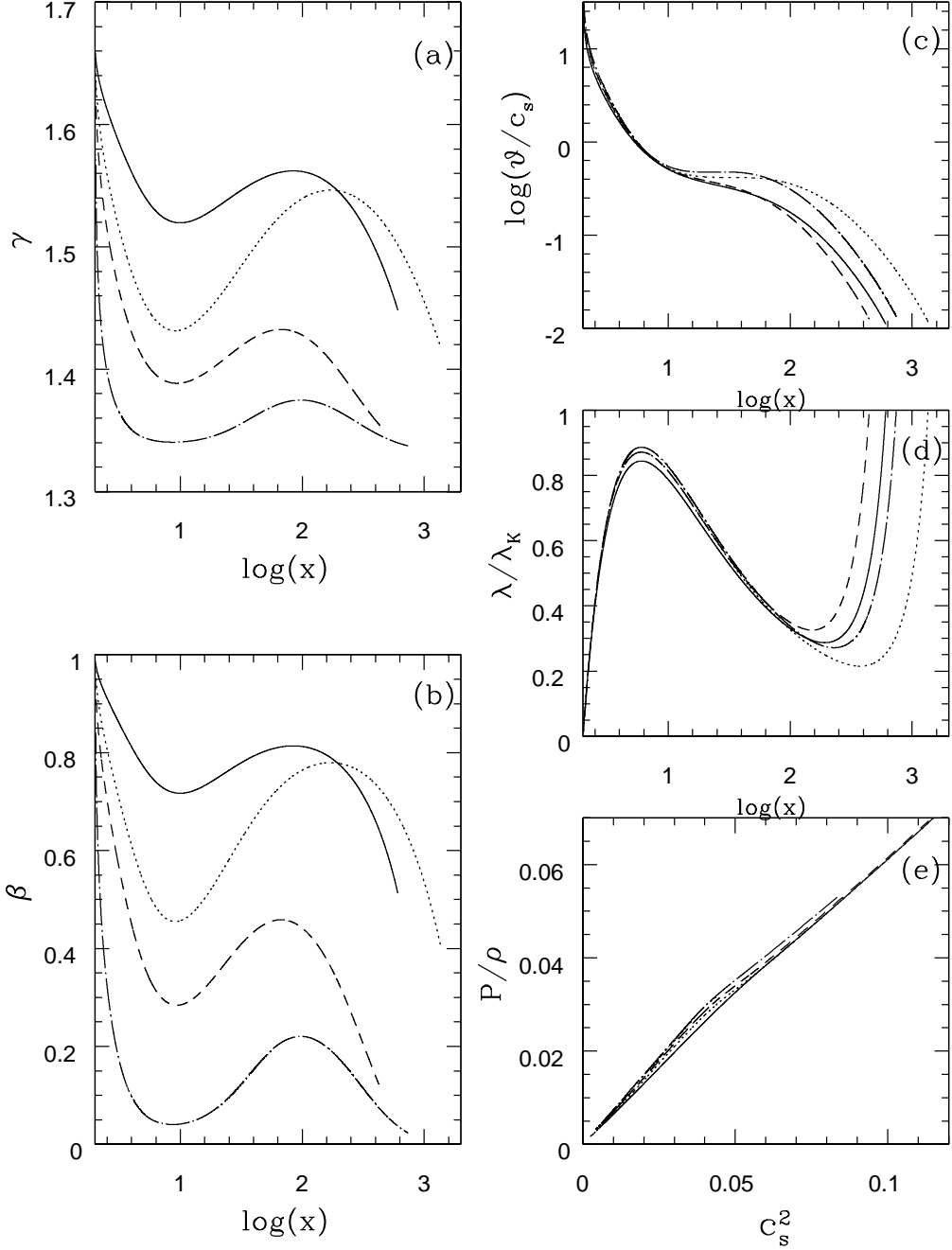}
   \caption{\small
   Variation of (a) polytropic index,
(b) ratio of gas pressure to total pressure, 
(c) Mach number, (d) ratio of specific angular momentum to 
the corresponding Keplerian value, (e) the density-pressure-sound speed relation (EOS). 
The solid, dotted, dashed, dot-long-dashed lines
corresponding to the different $\beta$ at transition radius are for $\beta_c=0.75,0.5,0.33,0.05$ respectively. 
Other parameters are $\alpha=0.01$, $f=0.5$, $x_c=5.8$, $a=0$, $M=10$. See Table 2 for detail.
}
              \label{figbr1}%
    \end{figure*}

There is a stronger variation of $\beta$ and then $\gamma$ in optically thin flows, compared to its
optically thick counter parts. As seen in Fig. \ref{figbr1}, while for a high $\beta_t$, 
the variation of $\beta$ could be $\sim 100\%$
and of $\gamma$ be $\sim 15\%$, for a low $\beta_t$, the respective variations are much higher,
$\sim 300\%$ and $\sim 20\%$. In all the cases depicted here, as $\lambda$
decreases around $x_t$ (see Fig. \ref{figbr1}d), rendering the flow to be quasi-spherical, $\beta$ increases as the
flow advances in. Subsequently, centrifugal force starts increasing, which renders the increase of $\lambda$ and
rate of cooling, which decreases $\beta$ and $\gamma$. However, close to the black hole, due to the dominance of
gravitational force, matter falls very fast, rendering the flows, irrespective of $\beta_t$,
to be quasi-spherical with high $\beta$. 
Therefore, unlike our conventional thought that highly relativistic flow in the vicinity of a black hole
may exhibit low $\beta$ and then $\gamma$ as seen in optically thick flows, in optically thin cases with
bremsstrahlung radiation
the situation is different. This is because the inefficient/incomplete cooling process (Rajesh \& Mukhopadhyay
2010a) keeps the flow to be very hot
and then quasi-spherical until inner edge when the infall time is very short.
It is very clear from EOSs shown in Fig. \ref{figbr1}e that $P/\rho$ does not vary linearly with $c_s^2$,
unlike the conventional cases. 

\begin{figure*}
   \centering
   \includegraphics[width=12cm]{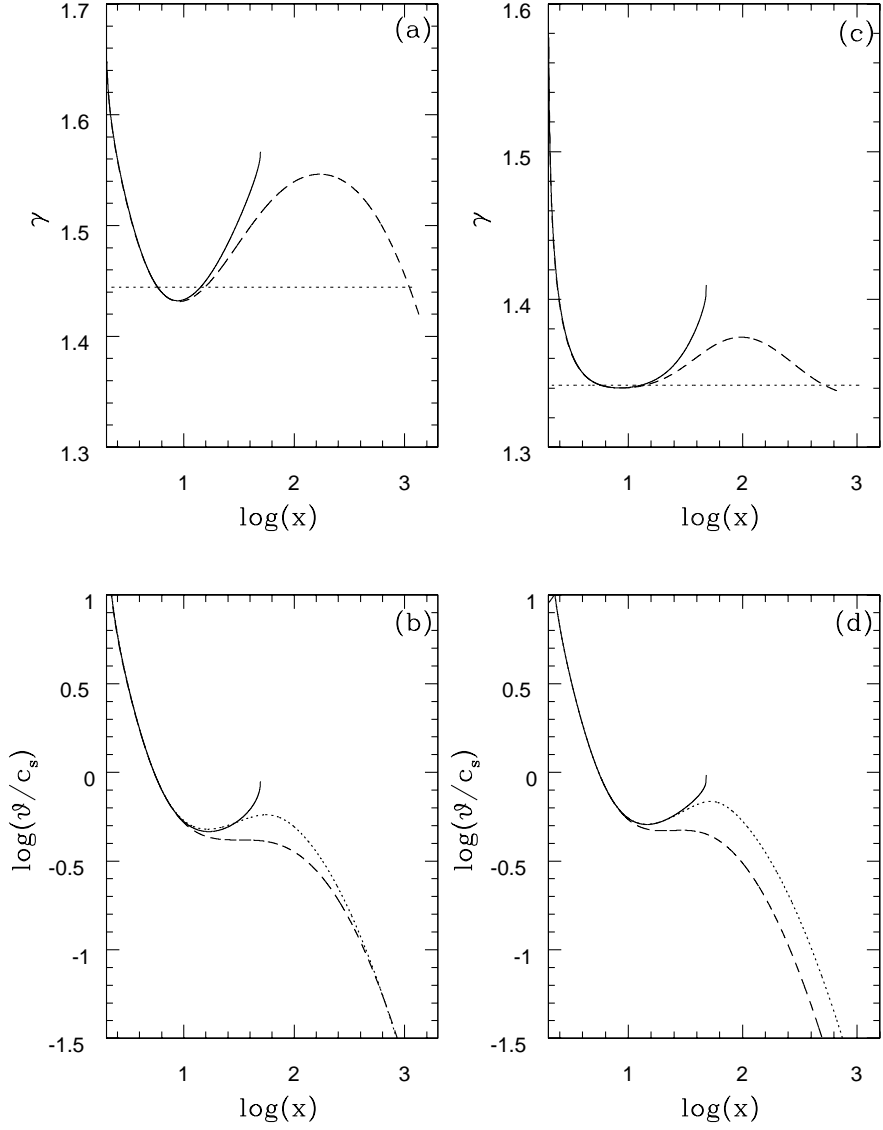}
   \caption{\small Variation of (a) polytropic index, (b) Mach number, as functions of radial coordinate.
The solid, dotted and dashed lines correspond to the flow with $\beta_c=0.5$, $\lambda_c=3.22$;
constant $\beta$, $\lambda_c=3.22$; and $\beta_c=0.5$, $\lambda_c=3.2$ respectively.
(c) and (d) are same as in (a) and (b) respectively, when the solid, dotted and dashed lines 
correspond to the flow with $\beta_c=0.05$, $\lambda_c=3.27$;
constant $\beta$, $\lambda_c=3.27$; and $\beta_c=0.05$, $\lambda_c=3.25$ respectively. Other
parameters are $\alpha=0.01$, $f=0.5$, $x_c=5.8$, $a=0$, $M=10$. See Table 2 for details.
}
              \label{figbr2}%
    \end{figure*}

Figure \ref{figbr2} shows that the self-consistent inclusion of variation of $\beta$ and then $\gamma$,
for both the gas and radiation dominated flows, 
renders the unphysical solutions (solid lines), when the corresponding solutions
with constant $\beta$ and then $\gamma$ (dotted lines) appear physical. It is very clear that at around $x=50$,
the value of $\gamma$ varies sharply with radius, rendering the flow to be gas dominated quasi-spherical
with an unphysical O-type critical radius. However, the flow with a smaller $\lambda_c$ (dashed lines),
keeping other parameters unchanged, exhibits the physical solution, even with varying $\beta$.
This is because the decrease of $\lambda$ favors the quasi-spherical nature of gas dominated flow at around
$x\gsim 50$, giving rise to a physical solution. In general, the specific angular momentum of the flow,
more specifically $\lambda_c$, is smaller in a realistic flow with varying $\beta$ and $\gamma$
than that with the idealistic choice of constant $\beta$ and $\gamma$.

\begin{figure*}
   \centering
   \includegraphics[width=14cm]{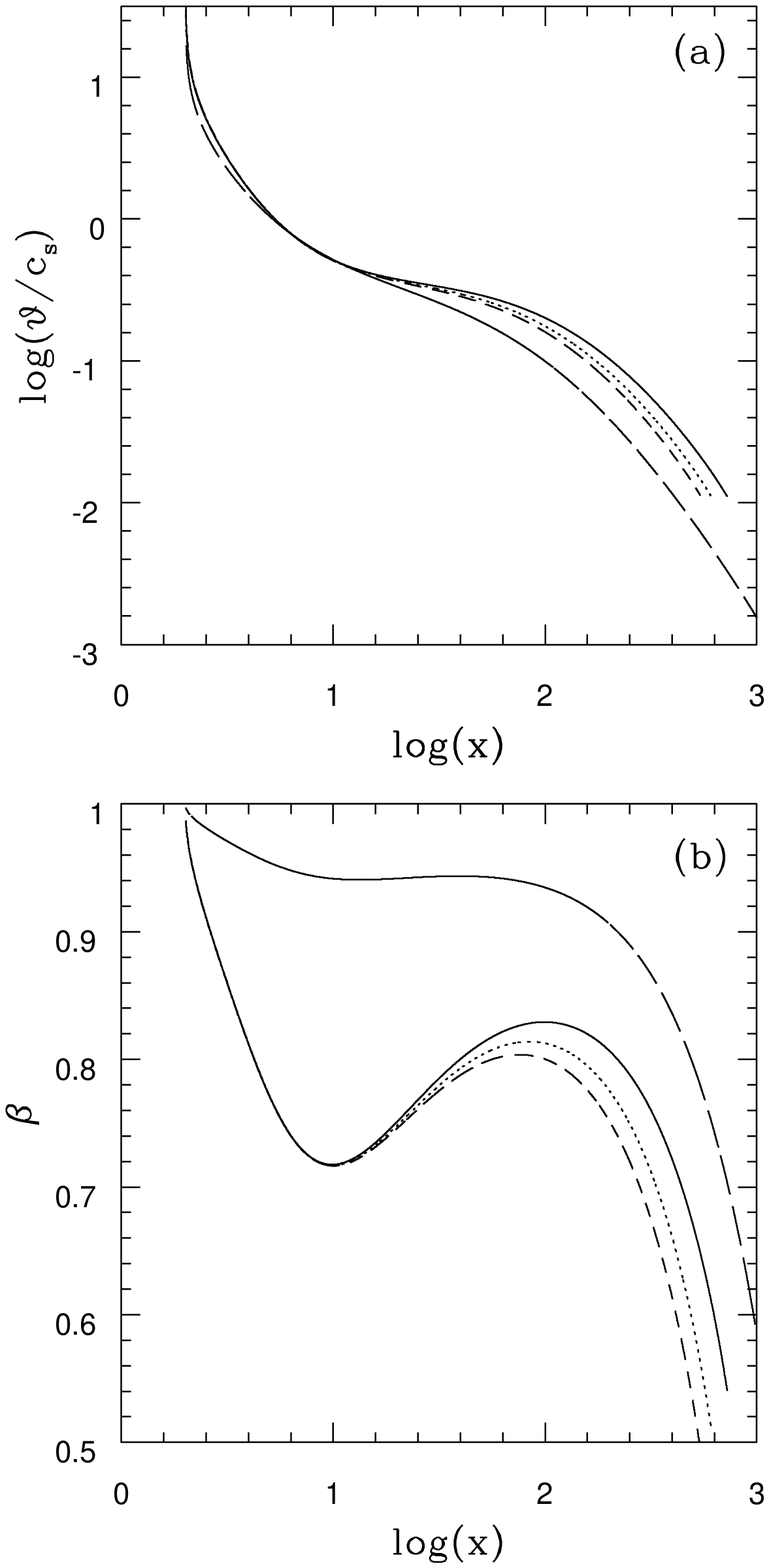}
   \caption{\small Variation of (a) Mach number, (b) ratio of gas pressure to total pressure, as functions of
of radial coordinate. The solid, dotted, dashed lines correspond to flows
with $f=0.1,0.5,0.8$ respectively when $\beta_c=0.75$, and the long-dashed line
corresponds to the flow with $f=0.98$ when $\beta_c=0.95$.
See Table 2 for detail.
}
              \label{figgascomf}%
    \end{figure*}

Figure \ref{figgascomf} shows that the variation of $f$ keeping other parameters
intact neither affects hydrodynamic properties nor $\beta$ profiles. However,
at a very high $f$ ($\rightarrow 1$, advection dominated flow), other parameters should not be
kept same and $\beta$ must be larger. Such a flow is represented by long-dashed lines. 
However, this still does not alter the Mach number profile significantly.
Note also that the flow with $\beta=f=1$ strictly at a large distance does not exhibit any 
variation of $\beta$ even at a smaller radius, which renders the corresponding 
$\gamma=5/3$, to be ratio of specific heats throughout. Therefore a strict ADAF solution 
remains unaltered. However, Narayan \& Yi (1994) in their ADAF chose $\gamma=1.5$, while
$\gamma=5/3$ corresponds to a Bondi flow. Moreover, for all the practical flows $\beta, f <1$ and hence
a variation of $\beta$ should be accounted in the model, in particular to explain luminosity,
as the present work intends to do.

%\clearpage
\begin{table}
\small
\caption{Parameter sets used for optically thick solutions: $M=10$, $\dot{m}$ is the accretion rate
in Eddington units}             % title of Table
\label{table1:1}      % is used to refer this table in the text
\centering                          % used for centering table
\begin{tabular}{c c c c c c c c}        % centered columns (4 columns)
\hline\hline                 % inserts double horizontal lines
$\beta_c$ & $\gamma_c$ & $\lambda_c$ & $f$ & $x_c$ & $a$ & $\alpha$ & $\dot{m}$\\    % table heading 
\hline                        % inserts single horizontal line
 $0.75$ & $1.533$ & $3.1$ & $0.5$ & $5.8$ & $0$ & $0.01$ & $0.01$ \\      % inserting body of the table
 $0.75$ & $1.533$ & $3.1$ & $0.5$ & $5.8$ & $0$ & $0.0001$ & $0.01$ \\      % inserting body of the table
 $0.75$ & $1.533$ & $3.05$ & $0.5$ & $5.8$ & $0$ & $0.01$ & $0.01$ \\      % inserting body of the table
 $0.75$ & $1.533$ & $3.05$ & $0.5$ & $5.8$ & $0$ & $0.0001$ & $0.01$ \\      % inserting body of the table
 $0.745$ & $1.531$ & $3.1$ & $0.5$ & $5.8$ & $0$ & $0.01$ & $0.01$ \\      % inserting body of the table
 $0.745$ & $1.531$ & $3.1$ & $0.5$ & $5.8$ & $0$ & $0.0001$ & $0.01$ \\      % inserting body of the table
 $0.7$ & $1.513$ & $3.1$ & $0.5$ & $5.8$ & $0$ & $0.0001$ & $0.01$ \\      % inserting body of the table
\hline                        % inserts single horizontal line
 $0.75$ & $1.533$  & $1.7$ & $0.5$ & $3.5$ & $0.998$ & $0.01$ & $0.01$ \\      % inserting body of the table
 $0.75$ & $1.533$ & $4$ & $0.5$ & $7.5$ & $-0.998$ & $0.01$ & $0.01$ \\      % inserting body of the table
 $0.75$ & $1.533$ & $3.95$ & $0.5$ & $7.5$ & $-0.998$ & $0.01$ & $0.01$ \\      % inserting body of the table
\hline                        % inserts single horizontal line
\hline                        % inserts single horizontal line
 $0.5$  & $1.444$ & $3.22$ & $0.5$ & $5.8$ & $0$ & $0.01$ & $0.01$\\
 $0.5$  & $1.444$ & $3.22$ & $0.5$ & $5.8$ & $0$ & $0.01$ & $0.01$\\
\hline                        % inserts single horizontal line
 $0.33$ & $1.399$ & $3.2$  & $0.1$ & $5.8$ & $0$ & $0.01$ & $1$ \\
 $0.33$ & $1.399$ & $3.2$  & $0.3$ & $5.8$ & $0$ & $0.01$ & $1$ \\
 $0.33$ & $1.399$ & $3.2$  & $0.7$ & $5.8$ & $0$ & $0.01$ & $1$ \\
 $0.33$ & $1.399$ & $3.2$  & $0.9$ & $5.8$ & $0$ & $0.01$ & $1$ \\
\hline                        % inserts single horizontal line
 $0.05$ & $1.342$ & $3.2$  & $0.3$ & $5.8$ & $0$ & $0.01$ & $1$ \\
 $0.05$ & $1.342$ & $1.8$  & $0.5$ & $3.5$ & $0.998$ & $0.01$ & $1$ \\
\hline                                   %inserts single line
\hline                        % inserts single horizontal line
& & & constant $\beta/\gamma$ & & &  \\
\hline                                   %inserts single line
 $0.75$ & $1.533$ & $3.1$ & $0.5$ & $5.8$ & $0$ & $0.01$ & $0.01$ \\      % inserting body of the table
 $0.75$ & $1.533$ & $3.1$ & $0.5$ & $5.8$ & $0$ & $0.0001$ & $0.01$ \\      % inserting body of the table
 $0.75$ & $1.533$ & $1.7$ & $0.5$ & $3.5$ & $0.998$ & $0.01$ & $0.01$ \\      % inserting body of the table
 $0.75$ & $1.533$ & $4$ & $0.5$ & $7.5$ & $-0.998$ & $0.01$ & $0.01$ \\      % inserting body of the table
\hline                                   %inserts single line
\hline                        % inserts single horizontal line
\end{tabular}
\end{table}

\begin{table}
\small
\caption{Parameter sets used for optically thin solutions: $x_c=5.8$, $a=0$, $\alpha=0.01$ 
(except for the case with $\beta_c=0.95$, when $\alpha=0.001$), $M=10$, 
$\dot{m}$ is the accretion rate in Eddington units}             % title of Table
\label{table1:1}      % is used to refer this table in the text
\centering                          % used for centering table
\begin{tabular}{c c c c c}        % centered columns (4 columns)
\hline\hline                 % inserts double horizontal lines
$\beta_c$ & $\gamma_c$ & $\lambda_c$ &  $f$ & $\dot{m}$\\    % table heading 
\hline                        % inserts single horizontal line
 $0.75$ & $1.533$ & $3.14$ & $0.5$ & $0.01$ \\      % inserting body of the table
 $0.75$ & $1.533$ & $3.1$ & $0.5$ & $0.01$ \\      % inserting body of the table
\hline                        % inserts single horizontal line
 $0.75$ & $1.533$ & $3.1$ & $0.1$ & $0.01$ \\      % inserting body of the table
 $0.75$ & $1.533$ & $3.1$ & $0.8$ & $0.01$ \\      % inserting body of the table
 $0.95$ & $1.635$ & $2.8$ & $0.98$ & $0.01$ \\      % inserting body of the table
\hline                        % inserts single horizontal line
\hline                        % inserts single horizontal line
 $0.5$  & $1.444$ & $3.22$ & $0.5$ & $0.01$\\
 $0.5$  & $1.444$ & $3.2$ & $0.5$ & $0.01$\\
\hline                        % inserts single horizontal line
\hline                        % inserts single horizontal line
 $0.33$ & $1.399$ & $3.25$ & $0.5$ & $1$ \\
 $0.33$ & $1.399$ & $3.2$  & $0.5$ & $1$ \\
\hline                        % inserts single horizontal line
\hline                        % inserts single horizontal line
 $0.05$ & $1.342$ & $3.27$ & $0.5$  & $1$ \\
 $0.05$ & $1.342$ & $3.25$  & $0.5$ & $1$ \\
\hline                                   %inserts single line
\hline                        % inserts single horizontal line
 & & constant $\beta/\gamma$  &  \\
\hline                                   %inserts single line
 $0.75$ & $1.533$ & $3.14$ & $0.5$ & $0.01$ \\      % inserting body of the table
 $0.5$ & $1.444$ & $3.22$ & $0.5$ & $0.01$ \\      % inserting body of the table
 $0.33$ & $1.399$ & $3.25$ & $0.5$ & $1$ \\      % inserting body of the table
 $0.05$ & $1.342$ & $3.27$ & $0.5$ & $1$ \\      % inserting body of the table
\hline                                   %inserts single line
\hline                        % inserts single horizontal line
\end{tabular}
\end{table}

%                                     Two column figure (place early!)
%______________________________________________ Gamma_1 (lg rho, lg e)

\section{Summary}

We have investigated the sub-Keplerian general 
advective accretion flows (GAAF; Rajesh \& Mukhopadhyay 2010a) around 
black holes allowing evolution of the gas and radiation content
into flows self-consistently. We have, in one hand, considered
radiation to be optically thick blackbody in the geometrically thin/slim vertically integrated disks.
On the other hand, we have also considered optically thin geometrically thicker flows
in the assumption of bremsstrahlung radiation.
Hence, as matter advances towards
a black hole, the ratio of the gas pressure to total pressure and the corresponding
polytropic index vary significantly with the radial coordinate. 
%This is, to best of our knowledge, a first investigation of accretion flows showing
%variation of $\beta$ and $\gamma$. 
We have found that in several occasions,
the accretion solutions as obtained 
with a constant $\beta$ by the earlier models do not exist when 
$\beta$ is allowed to vary self-consistently. It has a very important
implication when most of the existing models of accretion disks assume
$\beta$ and $\gamma$ to be constant throughout the flow. For example,
an advection dominated class of solution, namely ADAF (Narayan \& Yi 1994, 1995;
Quataert \& Narayan 1999),
has been modeled based on gas dominated flows with a constant $\gamma$ ($=1.5$) 
throughout, when $\gamma$ is chosen to be ratio of specific heats. 
Other models (e.g. Abramowicz et al. 1995; Chakrabarti 1996; 
Rajesh \& Mukhopadhyay 2010a),
describing gas dominated optically thin flows, as well have assumed $\gamma$, which is 
polytropic index therein,
to be constant. In the present work, the impact of varying $\gamma$ has been
particularly demonstrated in the
framework of later model. Note, however, that polytropic index turns out to be a ratio of specific 
heats only when the flow is of pure gas with $\gamma=5/3$. The results argue for correction to the earlier
models based on constant $\beta$, when particularly $\beta$ varies more than
$100\%$ during the infall of an optically thin gas dominated matter.
Hence, the present results imply that the entire branch of optically thin hot solutions of 
accretion disk needs to be re-investigated including appropriate variation of $\gamma$ and $\beta$.
However, for simplicity we have considered the optically thin radiation arising
due to bremsstrahlung mechanism only, neglecting synchrotron, inverse-Comptonization which
are likely to operate in the inner hot regime of the flow.
However, inclusion of such effects would only favor the present results.   
In future, we have plan to include such physics in studying very hot flows.
This will have immediate influence to the computed spectra (e.g. Yuan et al. 2003;
Mandal \& Chakrabarti 2005) and subsequent inferences about sources.

\section*{Acknowledgments}
This work is partly supported by a project, Grant No. SR/S2HEP12/2007, funded
by DST, India. One of the authors (PD) thanks the KVPY, DST, India, 
for providing a fellowship.

\end{document}